\documentstyle[aps,prl,epsfig]{revtex}
\begin{document}
%\draft command makes pacs numbers print
\draft
\twocolumn[\hsize\textwidth\columnwidth\hsize\csname %
@twocolumnfalse\endcsname

\title{ Bipolarons in the Extended Holstein Hubbard Model}
\author{ $^a$J. Bon\v ca and  $^b$S. A. Trugman}
\address{$^a$ FMF, University of Ljubljana and 
J. Stefan Institute, 1000, Slovenia,\\
$^b$Theory Division,
Los Alamos National Laboratory, Los Alamos, NM  87545,}
\date{\today}
\maketitle
\begin{abstract}\widetext

We numerically and analytically calculate
the properties of the bipolaron in an extended Hubbard
Holstein model, which has a longer range electron-phonon coupling
like the Fr\" ohlich model.
%We solve the  model with an exact
%diagonalization method on an infinite lattice and compare numerical results
%with analytical strong coupling calculations.
In the strong coupling
regime, the effective mass of the bipolaron in the extended model is much
smaller than the Holstein bipolaron mass.  In contrast to the Holstein
bipolaron, the bipolaron in the extended model has a lower binding energy and
remains bound with substantial binding energy even in the large-U
limit.
In comparison with the 
Holstein model where only a singlet bipolaron is bound, in the extended
Holstein model  a triplet bipolaron can also form a bound state. We discuss the
possibility of phase separation in the case of finite electron doping.

\end{abstract}

]

\narrowtext

There is growing evidence that electron-phonon coupling plays an
important role in determining exotic properties of novel materials
such as colossal magnetoresistance \cite{jin} and
high-$T_c$ compounds \cite{bednorz}. 
Since electrons in these
materials are strongly correlated, the interplay between an attractive
electron phonon interaction and Coulomb repulsion may be important in
determining physics at finite doping. In particular, when the
electron-phonon interaction is local, as is the case in the Holstein
model, finite Coulomb repulsion leads to the formation of an intra-site
bipolaron \cite{proville,fehske0,bonca1}, with an effective mass of the order
of the polaron effective mass \cite{bonca1}.

\begin{figure}[tb]
\begin{center}
\epsfig{file=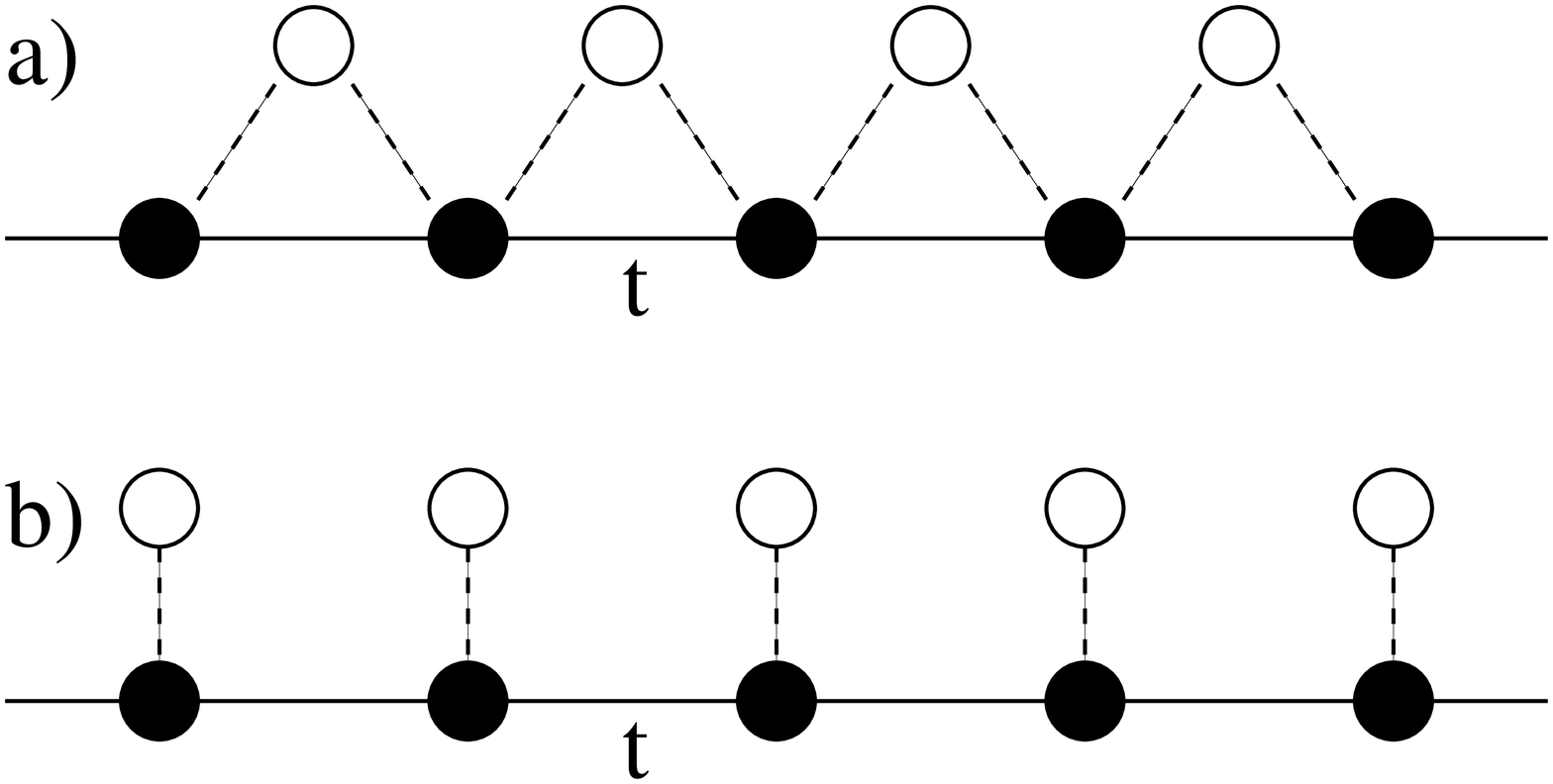,height=30mm,angle=-0}
\end{center}
\caption{ Schematic representation of the simplified a) extended
Holstein 
and b) Holstein model on a chain. 
Filled circles represent electron Wannier orbitals, open circles
represent ions. Solid lines indicate overlap integral $t$ between
Wannier orbitals, dashed lines represent nonzero electron-phonon
coupling.}
\label{model}
\end{figure}

It has been recently discovered that a
longer-range electron-phonon interaction leads to a decrease in the
effective mass of a polaron in the strong-coupling regime \cite{alex,fehske}.
The lower mass can have important
consequences, because lighter polarons and bipolarons
are more likely to remain mobile, and
less likely to trap on impurities or from mutual repulsion.
Motivated by this discovery, we investigate a simplified version of the
Fr\" ohlich model in the case of two electrons,
\begin{eqnarray}
H  = &-&t \sum_{js} ( c_{j+1,s}^\dagger c_{j,s} + H.c.)  \label{ham}\\
&-&\omega g_0\sum_{jls} f_l(j) c_{j,s}^\dagger c_{j,s}
( a_l + a_l^\dagger)  \nonumber \\
&+& \omega  \sum_j a_j^\dagger a_j + 
U\sum_{j}n_{j\uparrow}n_{j\downarrow},\nonumber
\end{eqnarray}
where $c_{j,s}^\dagger$ creates an electron of spin $s$ and
$a_{j}^\dagger$ creates a phonon on site $j$.  The second term
represents the coupling of an electron on site $j$ with an ion on site
$l$, where $g_0$ is the dimensionless electron-phonon coupling
constant.  While in general long range electron-phonon coupling
$f_l(j)$ is considered \cite{alex,fehske}, we further simplify this
model by placing ions in the interstitial sites located between
Wannier orbitals, as occurs in certain oxides \cite{tsuda},
shown in Fig.~(\ref{model}a).  In this case it is
natural to investigate a simplified model, where an electron located
on site $j$ couples only to its two neighboring ions, {\it i.e.} $l=j\pm 1/2$.
We describe such coupling with $f_{j\pm1/2}(j)=1$ and 0 otherwise,
and refer to this model as the extended Holstein Hubbard model (EHHM).
We can view the EHHM as the simplest model with longer range than a single
site, and use it to explore the qualitative change in physics in the simplest
possible setting.  While it is clear that in comparison to the Fr\"
ohlich model, our simplified EHHM lacks long range tails in the
electron phonon interaction, the physical properties that depend
predominantly on the short range interaction should be similar.
For example, calculating the polaron energy of the original Fr\" ohlich
model as defined in Refs.~\cite{alex,fehske}, one finds that 94\%
of the total polaron energy comes from the first two sites.

  In the case when $f_l(j)=\delta_{l,j}$, the model in
Eq.~(\ref{ham}) maps onto a Holstein-Hubbard model (HHM) (see also
Fig.(\ref{model}b)).  The last two terms in Eq.~(\ref{ham}) represent
the energy of the Einstein oscillator with frequency $\omega$ and the
on-site Coulomb repulsion between two electrons.  We consider the case
where two electrons with opposite spins ($S_z = 0$) couple to dispersionless
optical phonons with polarization perpendicular to the chain.

In this Letter we use a recently developed,
highly accurate numerical technique
\cite{bonca,bonca1}, combined with a strong coupling expansion to
study the simplified EHHM. Our main goal is to calculate 
physical properties such as the binding energy, effective mass,
isotope effect,
and the phase diagram of the EHHM bipolaron 
and compare them to the Holstein bipolaron
that has been thoroughly studied recently
\cite{bonca1}. Even though the two models appear very similar, we find
profound differences between the physical properties of bipolarons
within the EHHM and the HHM. 

The numerical method that we use creates a systematically
expandable variational space of phonon excitations
in the vicinity of the two electrons \cite{bonca,bonca1}. 
The variational method is defined on an infinite lattice and
is not subject to finite-size effects. It allows the calculation of
physical properties at any wavevector $k$. In the intermediate
coupling regime where it is most accurate, it provides results that are
variational in the thermodynamic limit and gives energies accurate to
14 digits for the polaron case and up to 7 digits for the bipolaron
case.

To investigate the strong coupling regime of the EHHM, we use a Lang-Firsov
\cite{lang} unitary transformation $\tilde H = e^S H e^{-S}$, where
$S = g_0\sum_{jls} f_l(j) n_{js}(a_l-a_l^\dagger)$.
This incorporates the exact distortion and interaction
energies for static electrons into $H_0$, and
leads to a transformed Hamiltonian
\begin{eqnarray}
\tilde H &=& H_0 + T,  \label{hamtilde}\\
H_0&=&\omega\sum_j  a_j^\dagger a_j 
-\omega g_0^2\sum_{ijl}f_l(i)f_l(j)n_i n_j
+ U\sum_{j}n_{j\uparrow}n_{j\downarrow} , \nonumber \\  
%H_0&=&\omega\sum_j  
%a_j^\dagger a_j - \omega g_0^2 \sum_{jl} f_l(j)^2 n_j \nonumber +
%U\sum_{j}n_{j\uparrow}n_{j\downarrow}\nonumber \\ 
%&-&2\omega g_0^2\sum_{jl}f_l(j)^2 n_{j\uparrow}n_{j\downarrow}\nonumber 
%-\omega g_0^2\sum_{i\not =jl}f_l(i)f_l(j)n_i n_j ,\\
%
T &=& -t e^{-\tilde g^2}\sum_{js} c_{j+1,s}^\dagger c_{j,s} 
e^{-g_0\sum_l\left(f_l(j+1) - f_l(j)\right)a_l^\dagger} \nonumber \\
&&e^{ g_0\sum_l\left(f_l(j+1) - f_l(j)\right)a_l} 
+ {\rm H.c.},\nonumber
\end{eqnarray}
where  $n_{j} = n_{j \uparrow} + n_{j \downarrow}$ and 
%
%\begin{equation}
$\tilde g^2 = g_0^2\sum_l [ f_l(0)^2-f_l(0)f_l(1) ];$
%\label{gtilde}
%\end{equation}
%
$ \tilde g = g_0$ for the EHHM.
The second term in $H_0$ gives the polaron energy, which in the EHHM case
is $\epsilon_p=2\omega g_0^2$, while for the HHM, $\epsilon_p=\omega
g_0^2$. 
This term also includes the interaction between electrons
located on neighboring sites, a consequence of the non-local
electron-phonon interaction.
As noted by Alexandrov and Kornilovitch \cite{alex}, in the
strong coupling regime a Fr\" ohlich polaron has a much smaller
effective mass than a Holstein polaron 
with the same polaron energy $\epsilon_p$.
The reason for lower mass in the Fr\" ohlich case (as well as EHHM) 
is that the effective electron-phonon
coupling that renormalizes hopping $\tilde g^2=\gamma
\epsilon_p/\omega$ is smaller  (in EHHM, $\gamma
= 1/2$) than in the case of the HHM with
$\gamma=1$.  
%The mass enhancement is caused by a mismatch in
%the phonon displacements in the initial and final states.
In the strong coupling EHHM polaron, the phonon 
is displaced on two sites.  It is identical
on one of these sites in the initial and the final state
after the electron hop, resulting in a smaller mass enhancement
from phonon overlap.

In the anti-adiabatic limit where $g_0\to 0$ and $\omega\to\infty$ with
$\omega g_0^2$ constant, the phonon interaction is instantaneous and
our simplified EHHM model maps onto a generalized Hubbard model
\begin{eqnarray}
H  &=& -t \sum_{js} ( c_{j+1,s}^\dagger c_{j,s} + H.c.) \nonumber \\
&+&\tilde U\sum_{j}n_{j\uparrow}n_{j\downarrow}
+V\sum_{j}n_{j}n_{j+1},
\label{hubb}
\end{eqnarray}
with an effective Hubbard interaction $\tilde U = U-4\omega g_0^2$ and
$V=-2\omega g_0^2$. In the case of two electrons an analytical solution
can be found. 
%In the general case when the two electrons have opposite spins, 
As many as three bound states may exist: two singlets and a triplet.
In the case when $U=0$ there is always at least one singlet
bound state. A triplet bound state with an energy $E=-2\omega
g_0^2-2t^2/\omega g_0^2$ exists only when $\omega g^2_0 > t$.

In the strong coupling limit, $T$ in Eq.~(\ref{hamtilde}) may be
considered as a perturbation.  In the case when $U<2\omega g_0^2$, the
single site or S0 bipolaron, defined as
$\phi_{S0}=c^\dagger_{0\uparrow}c^\dagger_{0\downarrow}\vert0\rangle$,
has the lowest energy to zeroth order. In this regime the binding
energy is $\Delta=E_{bi}^{S0}-2\epsilon_p=U-4\omega g_0^2$, where
$E_{bi}^{S0}$ denotes the S0 bipolaron energy and $\epsilon_p$ is the
energy of a polaron in zeroth order. 
%The binding at small $U$ is
%mainly due to the fourth term in $H_0$. 
In the opposite regime, when
$U>2\omega g_0^2$, the inter-site or S1 bipolaron,
$\phi_{S1}^{S=0,1}={1\over \sqrt 2}
(c^\dagger_{0\uparrow}c^\dagger_{1\downarrow}\pm
c^\dagger_{0\downarrow} c^\dagger_{1\uparrow})\vert0\rangle$, has the
lowest energy. Its binding energy $\Delta=-2\omega g_0^2$ does not
depend on $U$, which also leads to a degeneracy between the
spin-singlet (S=0) and the spin-triplet (S=1)  state.  
%In this case, binding
%is achieved due to the fifth term in $H_0$ which is absent in the HHM.
This simple analysis predicts that a EHHM bipolaron (EHB)
remains bound in the strong coupling regime even in the limit when
$U\to\infty$.

It is worth stressing that in the limit $U\to\infty$, singlet and
triplet bipolarons become degenerate. We can therefore predict the
existence of a singlet and a triplet bipolaron, where at finite $U$ the
singlet bipolaron has lower energy. It is also obvious that
the energy of the triplet bipolaron should not depend on $U$.  In
contrast to these predictions, a triplet Holstein bipolaron (HB) is
never stable, and furthermore in the limit $U\to\infty$ no bound HB
exists \cite{bonca1}.

Next, we focus on the effective mass of the EHB in the strong coupling
regime.  First order perturbation theory does not lead to
energy corrections for the S0 EHB.  Second order perturbation theory gives
\begin{equation}
{m^*_{S0}}^{-1} = {4t^2e^{-2g^2}}
\sum_{n= 0}{(-2g^2)^n\over n!}{1\over 
\epsilon_p-U + n\omega},
\label{mbi}
\end{equation}
where $m^{* -1} \equiv d^2 E(k) / dk^2 $.
Equation (\ref{mbi}) is only valid in the limit when
$1/\lambda \equiv 2t/\epsilon_p\to 0$ and $U \ll \epsilon_p$. In the limit of
large $g$ and $U=0$, ${m^*_{S0}}\propto \exp(2\epsilon_p/\omega)$,
which should be compared to the HB effective mass that scales as
${m^*_{S0}}\propto \exp(4\epsilon_p/\omega)$ \cite{bonca1,kabanov}. In
the strong coupling regime the EHB  should be much 
lighter than the Holstein bipolaron.
There is a particularly interesting EHB regime when $U=\epsilon_p$.
In this case the zero order energies of the $\phi_{S0}$ and
$\phi_{S1}^{S=0,1}$ bipolarons are degenerate.  Degenerate first
order perturbation theory can be applied to the spin-singlet EHB
in this case, which leads to
a substantial decrease in the effective mass
\begin{equation}
m_{EHB}^*(U=\epsilon_p)=
{\sqrt 2\over t} e^{\epsilon_p/2\omega}.
\label{massf1}
\end{equation}
The EHB in this regime consists of a superposition of $\phi_{S0}$ and 
$\phi_{S1}^{S=0}$, and moves through the lattice in a crab-like
motion. Its binding energy is
$\Delta = -\epsilon_p - 2 \sqrt 2 t \exp(-\epsilon_p/2\omega)$.

In the $U\to\infty$ limit  we apply the second-order
perturbation theory to the S1 bipolaron. We take into account 
processes where one of the electrons within the S1 bipolaron jumps to
the left (right) and then the other follows. This leads to 
\begin{equation}
m_{EHB}^*(U=\infty)={\lambda\over t}e^{\epsilon_p/\omega}.
\label{massf2}
\end{equation}
Strong coupling approach thus predicts nonmonotonous dependence of the
effective EHB mass as a function of $U$ as can be seen from different
exponents in Eqs.~(\ref{mbi},\ref{massf1},\ref{massf2}).
\begin{figure}[tb]
\begin{center}
\epsfig{file=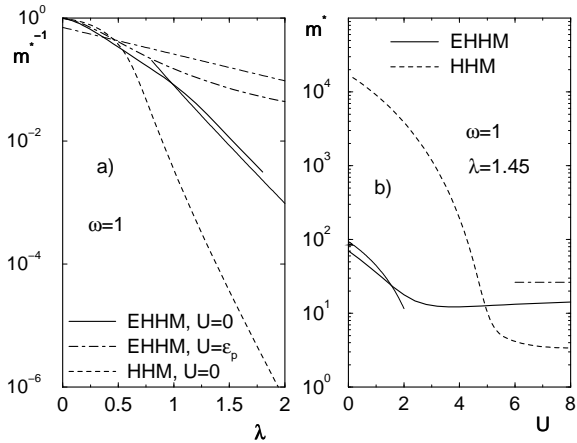,height=75mm,angle=-90}
\end{center}
\caption{ a) The bipolaron inverse effective mass vs. $\lambda$ at $\omega=1$.
The thin full line and thin dot-dashed line represent strong coupling
results obtained using Eqs.~(\ref{mbi}) and (\ref{massf1})
respectively.
b) The effective mass
vs. $U$ at $\omega=1$ and $\lambda=1.45$. The thin full and dot-dashed
lines represent strong coupling results obtained using
Eqs.~(\ref{mbi}) and (\ref{massf2}) respectively.  }
\label{fmass1}
\end{figure}
We next present numerical results. To achieve sufficient accuracy, we
have used up to $ 3~\times~10^6$ variational states.  We use units
where the bare hopping constant is $t=1$.  The ground state energy of
the  EHB at $\lambda = 0.5$, $\omega = U = 1$, is
E = -5.822621, which is accurate to the number of
digits shown.  (For the same parameters, $U=0$, the Holstein bipolaron
energy is E = ~ -5.4246528.)  The accuracy of our plotted results in the
thermodynamic limit is well within the line-thickness. In
Fig.~(\ref{fmass1}a) we present the inverse effective masses of the EHB
and the HB at $U=0$ and of the EHB at $U=\epsilon_p$. Our results for
the bipolaron mass are in qualitative agreement with results for the
polaron effective mass by Alexandrov and Kornilovitch \cite{alex}. In
the weak coupling regime we find the EHB slightly heavier than the HB,
while in the strong coupling regime the opposite is true. Setting the
Coulomb interaction to $U=\epsilon_p$, the effective mass becomes even
lighter, which is a consequence of the smaller exponent in
Eq.~(\ref{massf1}).  In the strong coupling regime ($\lambda \geq 1$),
we find good agreement with our strong coupling predictions 
in Eqs.~(\ref{mbi},\ref{massf1}), depicted by thin lines.
While the absolute values
may differ by up to a factor of 4 (in the case of $U=\epsilon_p$), the strong
coupling approach almost perfectly predicts the exponential
dependence (seen as parallel straight lines in Fig.~(\ref{fmass1}a)) of
the effective masses on $\epsilon_p = 2t \lambda $.

To obtain  better understanding of the effect of on-site Coulomb repulsion on
the bipolaron effective mass in the strong coupling regime, we present
in Fig.~(\ref{fmass1}b) effective masses of the EHB and HB at fixed
coupling strength $\lambda=1.45$ as a function of 
$U$. The most prominent finding is that the EHB is two
orders of magnitude lighter than the HB when $U=0$. While the
effective mass of the HB decreases monotonously with  $U$,
the EHB effective mass reaches a shallow minimum near $U=\epsilon_p$
as predicted by the strong coupling approach. At larger $U>\epsilon_p$
we observe a slight increase in the effective mass. In the same regime
HB effective mass drops below EHB effective mass. This crossing
coincides with a substantial decrease of the HB binding energy and
consequently with separating of HB into two separate polarons. 
Numerical results for the EHB agree reasonably well
with analytical predictions for small $U$, Eq.~(\ref{mbi}), and
also in the limit of large $U$, Eq.~(\ref{massf2}).

\begin{figure}[tb]
\begin{center}
\epsfig{file=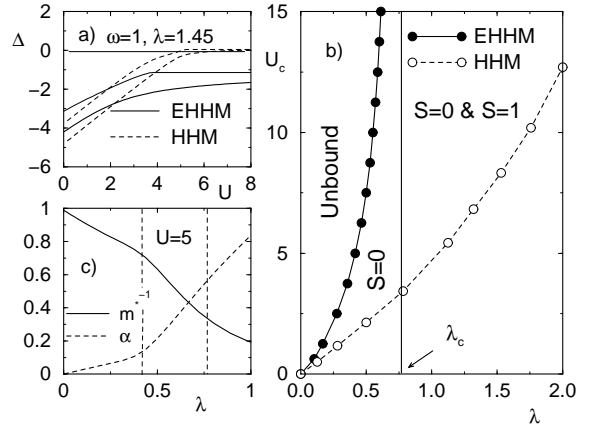,height=75mm,angle=-90}
\end{center}
\caption{ a) Binding energies $\Delta^{(0,1)}$ vs. $U$ of the EHHM
(full lines) and the HHM (dashed lines). Corresponding binding
energies of the first excited states are indicated with thin lines.
b) Phase diagram of the EHHM (filled circles) and the HHM
(open circles) calculated at $\omega =1$. The vertical line at
$\lambda=\lambda_c$ represents the stability line of the $S=1$ EHB. Text in
the figure applies only to EHHM phase diagram. c) The inverse effective
mass and the isotope effect of the EHHM vs. $\lambda$ at $U=5$. Vertical lines
represent stability limits of the $S=0$ and $S=1$ EHB (from left to
right). }
\label{fbind}
\end{figure}

To gain an insight into the symmetry of the bound EHB state, 
we have calculated the binding energy $\Delta^{(0,1)}=E_{bi}^{(0,1)}-2E_{po}$
where $E_{bi}^{(0,1)}$ are the ground state and the first excited
energy  of the EHHM or HHM for two electrons with opposite spins, $S_z=0$,
%(total spin $S=$ 0 and 1),
and $E_{po}$ is the ground state energy of the corresponding model with
one electron.  In Fig.~(\ref{fbind}a) we present binding energies of
the bipolaron ground and first excited states as a function of
$U$.
An important difference between the HHM and the EHHM is that in the
former case a critical $U_c$ exists for any coupling strength
$\lambda$ when the HB unbinds, while the EHB remains bound even in the
limit $U\to\infty$ when $\lambda>\lambda_c=0.76$.  At small $U$
excited states of both models correspond to bipolaronic singlets,
spaced approximately $\omega$ above the ground state. Singlets can be
recognized by the fact that their binding energies depend on $U$. As
$U$ increases, the excited state of the HB unbinds while the excited
state of the EHB undergoes a transition from a singlet to a triplet
state which is also  bound.

By solving $\Delta^{(0,1)}(\lambda,U_c)=0$ we arrive at the phase
diagram $(U_c,\lambda)$ of the EHHM 
calculated at fixed $\omega=1$,
presented in Fig.~(\ref{fbind}b). 
%of the Fr\"
%ohlich-Hubbard model calculated at fixed $\omega=1$.  
We indicate
three different regimes. For small $\lambda$ and large $U$ no bound
bipolarons exist. With increasing $\lambda$ there is a phase
transition into a bound singlet bipolaron state.  Increasing $\lambda$
even further, a triplet bipolaron becomes bound as well at
$\lambda=\lambda_c$.  For comparison we also include the phase
boundary of the HHM (open circles). Note that only a singlet bipolaron
exists in the HHM.

In Fig.~(\ref{fbind}c) we present a cross section through the phase
diagram in Fig.~(\ref{fbind}b) at fixed $U=5$, and plot
${m^*}^{-1}$ and the isotope effect $\alpha \equiv d \ln m_{bi}/ d \ln
M $ vs.  $\lambda$ (see also discussion of the isotope effect in
Ref.~\cite{bonca1}). The effective mass increases by approximately a
factor of 2.5 from its noninteracting value in the regime where only
a spin-singlet bipolaron exists (between the two vertical dashed
lines). The increase of the effective mass is followed by an increase
in the isotope effect. The binding energy (not plotted) reaches a
value $\Delta\sim -0.5 t$ at $\lambda=\lambda_c=0.76$.

To conclude, we have shown that a light EHB exists even in the strong
coupling regime with an effective mass that can be a few orders of
magnitude smaller than the HB effective mass at small $U$. At finite
$U=\epsilon_p$ a regime of extremely light EHB is found where
bipolaron effective mass scales with the same exponent as the polaron
effective mass. This mobile bipolaron
arises as a superposition of a $\phi_{S0}$ and a $\phi_{S1}$
state and it  moves through a lattice in a crab-like motion. 
As found in ref.~ \cite{bonca1}, HB becomes very
light with increasing $U$ close to the transition into
two unbound polarons at $U=U_c$. Near this transition, its binding
energy diminishes substantially and reaches $\Delta=0$ at the
transition point $U_c$.  In contrast, EHB can have a small
effective mass even in the regime where its  binding energy is large
( in the strong
coupling regime $\Delta$  approaches $\Delta = -\epsilon_p$).
Furthermore, EHB remains bound in the limit when $U\to\infty$. 
As a consequence of a longer-range electron-phonon interaction, a
bound spin-triplet bipolaron exists in the EHHM for $\lambda>\lambda_c$ 
The difference
between the binding energies of the spin-singlet and the spin-triplet
bipolaron is proportional to $1/U$.
In the weak to intermediate coupling regime of the EHHM
($\lambda<\lambda_c$ and finite $U$) $S=0$ bipolarons exist
with substantial binding
energy close to $\lambda\sim \lambda_c$,
and an effective mass of the order of noninteracting electron
mass.

The existence of a singlet and a triplet EHB state has important
implications in the case of finite doping. As was established
previously, there is no phase separation in the low-density limit of
the HHM despite a substantially renormalized bandwidth \cite{bonca1}.
The reason is in part that a triplet bipolaron is always unstable. The
lack of phase separation in the low-density limit and in the strong
coupling regime has a simple intuitive explanation: a third particle,
added to a bound singlet bipolaron, introduces a triplet component to
the wavefunction.  The opposite is true in the strong coupling limit
of EHHM where singlet and triplet bipolarons coexist. In this case,
the third added particle simply attaches to the existing singlet
bipolaron and thus gains in the potential energy.  We therefore expect
that the EHHM phase separates in the case of finite doping for
$\lambda$ sufficiently large.
% The system is unstable with respect to
%phase separation into a fully spin-polarized state, with unpolarized
%states even lower in energy.
To stabilize a system of EHHM
bipolarons against phase separation, a long-range Coulomb repulsion
should be taken into account. This prediction is in agreement with
recent findings by Alexandrov and Kabanov \cite{kabanov1} that state, 
that there is no phase
segregation in the Fr\" ohlich model in the presence of long-range Coulomb
interactions.

%As we have shown in our previous work
%\cite{bonca1}, HB also becomes very light with increasing $U$ close to
%transition into two unbound polarons $U=U_c$. At the same time the
%binding energy of the HB $\Delta\to 0$. In contrast, EHB remains bound
%even in the limit when $U\to \infty$ with binding energy that in the
%strong coupling regime approaches $\Delta = -\epsilon_p/2$. 

% I actually have results for the case of U(j)=U/(1+epsilon * abs(j))
% Introducing such U does not change much the effective mass.

J.B.  gratefully  acknowledges the    support of Los   Alamos National
Laboratory   where part  of  this   work has  been   performed, and
financial support by the Slovene Ministry of Science, Education and Sport.
This work was supported in part by the US DOE.

\end{document}